\input amstex
\documentstyle{amsppt}
%
\catcode`@=11
\redefine\output@{%
  \def\break{\penalty-\@M}\let\par\endgraf
  \ifodd\pageno\global\hoffset=105pt\else\global\hoffset=8pt\fi  
  \shipout\vbox{%
    \ifplain@
      \let\makeheadline\relax \let\makefootline\relax
    \else
      \iffirstpage@ \global\firstpage@false
        \let\rightheadline\frheadline
        \let\leftheadline\flheadline
      \else
        \ifrunheads@ 
        \else \let\makeheadline\relax
        \fi
      \fi
    \fi
    \makeheadline \pagebody \makefootline}%
  \advancepageno \ifnum\outputpenalty>-\@MM\else\dosupereject\fi
}
\catcode`\@=\active
\nopagenumbers
\def\negskp{\hskip -2pt}
\loadbold
\accentedsymbol\bulpsi{\overset{\kern 2pt\sssize\bullet\kern -2pt}%
   \to\psi}
\accentedsymbol\circpsi{\overset{\kern 2pt\sssize\circ\kern -2pt}%
   \to\psi}

\def\chirk{\special{em:point 1}\kern 1.2pt\raise 0.6pt
  \hbox to 0pt{\special{em:point 2}\hss}\kern -1.2pt
  \special{em:line 1,2,0.3pt}\ignorespaces}
\def\Chirk{\special{em:point 1}\kern 1.5pt\raise 0.6pt
  \hbox to 0pt{\special{em:point 2}\hss}\kern -1.5pt
  \special{em:line 1,2,0.3pt}\ignorespaces}
\def\Hirk{\kern 0pt\special{em:point 1}\kern 4pt\special{em:point 2}
  \kern -3pt\special{em:line 1,2,0.3pt}\ignorespaces}
\accentedsymbol\uuud{d\unskip\kern -3.8pt\raise 1pt\hbox to
  0pt{\chirk\hss}\kern 0.2pt\raise 1.9pt\hbox to 0pt{\chirk\hss}
  \kern 0.2pt\raise 2.8pt\hbox to 0pt{\chirk\hss}\kern 3.5pt}
\accentedsymbol\bolduuud{\bold d\unskip\kern -4pt\raise 1.1pt\hbox to
  0pt{\chirk\hss}\kern 0pt\raise 1.9pt\hbox to 0pt{\chirk\hss}
  \kern 0pt\raise 2.5pt\hbox to 0pt{\chirk\hss}\kern 4pt}
\accentedsymbol\uud{d\unskip\kern -3.72pt\raise 1.5pt\hbox to
  0pt{\chirk\hss}\kern 0.26pt\raise 2.4pt\hbox to 0pt{\chirk\hss}
  \kern 3.5pt}
\accentedsymbol\bolduud{\bold d\unskip\kern -4pt\raise 1.4pt\hbox to
  0pt{\chirk\hss}\kern 0pt\raise 2.3pt\hbox to 0pt{\chirk\hss}
  \kern 4pt}
\accentedsymbol\uuuD{D\unskip\kern -4.6pt\raise 2pt\hbox to
  0pt{\Chirk\hss}\kern 0.2pt\raise 3pt\hbox to 0pt{\Chirk\hss}
  \kern 0.2pt\raise 4.1pt\hbox to 0pt{\Chirk\hss}\kern 4.5pt}
\accentedsymbol\uuD{D\unskip\kern -4.5pt\raise 2.4pt\hbox to
  0pt{\Chirk\hss}\kern 0.2pt\raise 3.4pt\hbox to 0pt{\Chirk\hss}
  \kern 4.4pt}
\accentedsymbol\uD{D\unskip\kern -4.4pt\raise 3pt\hbox to
  0pt{\Chirk\hss}\kern 4.4pt}
\accentedsymbol\bolduuuD{\bold D\unskip\kern -4.6pt\raise 2pt\hbox to
  0pt{\Chirk\hss}\kern 0pt\raise 3pt\hbox to 0pt{\Chirk\hss}
  \kern 0pt\raise 4.1pt\hbox to 0pt{\Chirk\hss}\kern 4.5pt}
\accentedsymbol\bolduuD{\bold D\unskip\kern -4.5pt\raise 2.4pt\hbox to
  0pt{\Chirk\hss}\kern 0pt\raise 3.4pt\hbox to 0pt{\Chirk\hss}
  \kern 4.4pt}
\accentedsymbol\bolduD{\bold D\unskip\kern -4.5pt\raise 3pt\hbox to
  0pt{\Chirk\hss}\kern 4.4pt}
\accentedsymbol\uuuU{U\unskip\kern -5.1pt\raise 2pt\hbox to
  0pt{\Chirk\hss}\kern 0.2pt\raise 3pt\hbox to 0pt{\Chirk\hss}
  \kern 0.2pt\raise 4.1pt\hbox to 0pt{\Chirk\hss}\kern 4.9pt}
\accentedsymbol\uuU{U\unskip\kern -5.0pt\raise 2.4pt\hbox to
  0pt{\Chirk\hss}\kern 0.2pt\raise 3.4pt\hbox to 0pt{\Chirk\hss}
  \kern 4.9pt}
\accentedsymbol\uU{U\unskip\kern -4.9pt\raise 3pt\hbox to
  0pt{\Chirk\hss}\kern 4.8pt}
\accentedsymbol\bolduPsi{\boldsymbol\Psi\unskip\kern -6.5pt\raise 3pt
  \hbox to 0pt{\Hirk\hss}\kern 6.6pt}
\accentedsymbol\bolduuPsi{\boldsymbol\Psi\unskip\kern -6.5pt\raise 3pt
  \hbox to 0pt{\Hirk\hss}\unskip\kern 0pt\raise 3.8pt
  \hbox to 0pt{\Hirk\hss}\kern 6.6pt}
\accentedsymbol\bolduuuPsi{\boldsymbol\Psi\unskip\kern -6.5pt\raise 3pt
  \hbox to 0pt{\Hirk\hss}\unskip\kern 0pt\raise 3.8pt
  \hbox to 0pt{\Hirk\hss}\unskip\kern 0pt\raise 4.6pt
  \hbox to 0pt{\Hirk\hss}\kern 6.6pt}
\accentedsymbol\uA{\operatorname{A}\unskip\kern -6.2pt\raise 2.8pt\hbox to
  0pt{\vbox{\hrule width 2pt height 0.3pt}\hss}\kern 4.5pt}
\accentedsymbol\uuA{\operatorname{A}\unskip\kern -6.2pt\raise 2.8pt\hbox 
  to 0pt{\vbox{\hrule width 2pt height 0.3pt}\hss}\unskip\kern 0.2pt\raise
  3.4pt\hbox to 0pt{\vbox{\hrule width 1.6pt height 0.3pt}\hss}\kern 4.3pt}
\accentedsymbol\uuuA{\operatorname{A}\unskip\kern -6.2pt\raise 2.8pt\hbox 
  to 0pt{\vbox{\hrule width 2pt height 0.3pt}\hss}\unskip\kern 0.2pt\raise
  3.4pt\hbox to 0pt{\vbox{\hrule width 1.6pt height 0.3pt}\hss}\unskip
  \kern 0.2pt\raise 4.0pt\hbox to 0pt{\vbox{\hrule width 1.0pt height
  0.3pt}\hss}\kern 4.3pt}
\accentedsymbol\uF{\Cal F\unskip\kern -6.4pt\raise
  3.4pt\hbox to 0pt{\vbox{\hrule width 1.6pt height 0.3pt}\hss}\kern 6.7pt}
\accentedsymbol\uuF{\Cal F\unskip\kern -6.6pt\raise 2.8pt\hbox 
  to 0pt{\vbox{\hrule width 2pt height 0.3pt}\hss}\unskip\kern 0.2pt\raise
  3.4pt\hbox to 0pt{\vbox{\hrule width 1.6pt height 0.3pt}\hss}\kern 6.9pt}
\accentedsymbol\uuuF{\Cal F\unskip\kern -6.6pt\raise 2.8pt\hbox 
  to 0pt{\vbox{\hrule width 2pt height 0.3pt}\hss}\unskip\kern 0.2pt\raise
  3.4pt\hbox to 0pt{\vbox{\hrule width 1.6pt height 0.3pt}\hss}\unskip
  \kern 0.2pt\raise 4.0pt\hbox to 0pt{\vbox{\hrule width 1.45pt height
  0.3pt}\hss}\kern 6.9pt}
\accentedsymbol\uCA{\Cal A\unskip\kern -5.6pt\raise
  3.4pt\hbox to 0pt{\vbox{\hrule width 1.6pt height 0.3pt}\hss}\kern 5.9pt}
\accentedsymbol\uuCA{\Cal A\unskip\kern -5.9pt\raise 2.8pt\hbox 
  to 0pt{\vbox{\hrule width 2pt height 0.3pt}\hss}\unskip\kern 0.2pt\raise
  3.4pt\hbox to 0pt{\vbox{\hrule width 1.6pt height 0.3pt}\hss}\kern 6.2pt}
\accentedsymbol\uuuCA{\Cal A\unskip\kern -5.9pt\raise 2.8pt\hbox 
  to 0pt{\vbox{\hrule width 2pt height 0.3pt}\hss}\unskip\kern 0.3pt\raise
  3.4pt\hbox to 0pt{\vbox{\hrule width 1.6pt height 0.3pt}\hss}\unskip
  \kern 0.5pt\raise 4.0pt\hbox to 0pt{\vbox{\hrule width 1.45pt height
  0.3pt}\hss}\kern 6.2pt}
\loadeurb
\accentedsymbol\bolduF{\eurb F\unskip\kern -5.5pt\raise
  2.8pt\hbox to 0pt{\vbox{\hrule width 1.6pt height 0.3pt}\hss}\kern 5.3pt}
\accentedsymbol\bolduuF{\eurb F\unskip\kern -5.5pt\raise 2.8pt\hbox 
  to 0pt{\vbox{\hrule width 1.6pt height 0.3pt}\hss}\unskip\kern 0pt\raise 
  3.4pt\hbox to 0pt{\vbox{\hrule width 1.6pt height 0.3pt}\hss}\kern 5.3pt}
\accentedsymbol\bolduuuF{\eurb F\unskip\kern -5.5pt\raise 2.8pt\hbox 
  to 0pt{\vbox{\hrule width 1.6pt height 0.3pt}\hss}\unskip\kern 0pt\raise 
  3.4pt\hbox to 0pt{\vbox{\hrule width 1.6pt height 0.3pt}\hss}\unskip
  \kern 0pt\raise 4.0pt\hbox to 0pt{\vbox{\hrule width 1.6pt height
  0.3pt}\hss}\kern 5.3pt}
\accentedsymbol\bolduCA{\eurb A\unskip\kern -7.9pt\raise
  2.8pt\hbox to 0pt{\vbox{\hrule width 1.6pt height 0.3pt}\hss}\kern 7.7pt}
\accentedsymbol\bolduuCA{\eurb A\unskip\kern -7.9pt\raise 2.8pt\hbox 
  to 0pt{\vbox{\hrule width 1.6pt height 0.3pt}\hss}\unskip\kern 0.3pt
  \raise 3.4pt\hbox to 0pt{\vbox{\hrule width 1.6pt height 0.3pt}\hss}
  \kern 7.7pt}
\accentedsymbol\bolduuuCA{\eurb A\unskip\kern -7.9pt\raise 2.8pt\hbox 
  to 0pt{\vbox{\hrule width 1.6pt height 0.3pt}\hss}\unskip\kern 0.3pt
  \raise 3.4pt\hbox to 0pt{\vbox{\hrule width 1.6pt height 0.3pt}\hss}
  \unskip\kern 0.3pt\raise 4.0pt\hbox to 0pt{\vbox{\hrule width 1.6pt 
  height 0.3pt}\hss}\kern 7.7pt}
\def\blue#1{#1}
\catcode`#=11\def\diez{#}\catcode`#=6
\catcode`_=11\def\podcherkivanie{_}\catcode`_=8
\def\mycite#1{\cite{\blue{#1}}\immediate\special{ps:
     ShrHPSdict begin /ShrBORDERthickness 0 def}}
\def\myciterange#1#2#3#4{\cite{\blue{#2#3#4}}\immediate\special{ps:
     ShrHPSdict begin /ShrBORDERthickness 0 def}}
\def\mytag#1{%
    \tag#1}
\def\mythetag#1{\thetag{\blue{#1}}\immediate\special{ps:
     ShrHPSdict begin /ShrBORDERthickness 0 def}}
\def\myrefno#1{\no#1}
\def\myhref#1#2{\blue{#2}\immediate\special{ps:
     ShrHPSdict begin /ShrBORDERthickness 0 def}}
\def\myEarXivlink{\myhref{http://arXiv.org}{http:/\negskp/arXiv.org}}

\def\mytheorem#1{\csname proclaim\endcsname{Theorem #1}}

\def\mylemma#1{\csname proclaim\endcsname{Lemma #1}}

\def\mycorollary#1{\csname proclaim\endcsname{Corollary #1}}

\pagewidth{360pt}
\pageheight{606pt}
\topmatter
\title
The Higgs field can be expressed\\
through the lepton and quark fields.
\endtitle
\author
R.~A.~Sharipov
\endauthor
\address 5 Rabochaya street, 450003 Ufa, Russia\newline
\vphantom{a}\kern 12pt Cell Phone: +7(917)476-93-48
\endaddress
\email \vtop to 30pt{\hsize=280pt\noindent
\myhref{mailto:r-sharipov\@mail.ru}
{r-sharipov\@mail.ru}\newline
\myhref{mailto:R\podcherkivanie Sharipov\@ic.bashedu.ru}
{R\_\hskip 1pt Sharipov\@ic.bashedu.ru}\vss}
\endemail
\urladdr
\vtop to 20pt{\hsize=280pt\noindent
\myhref{http://www.geocities.com/r-sharipov/}
{http:/\negskp/www.geocities.com/r-sharipov}\newline
\myhref{http://www.freetextbooks.boom.ru/}
{http:/\negskp/www.freetextbooks.boom.ru}\vss}
\endurladdr
\abstract
    The Higgs field is a central point of the Standard Model 
supplying masses to other fields through the symmetry breaking 
mechanism. However, it is associated with an elementary particle 
which is not yet discovered experimentally. In this short note 
I suggest a way for expressing the Higgs field through other 
fields of the Standard Model. If this is the case, being not an
independent field, the Higgs field does not require an 
elementary particle to be associated with it.
\endabstract
\endtopmatter
\loadeufb
\TagsOnRight
\document

\rightheadtext{The Higgs field can be expressed through \dots}
\head
1. Matter fields of the Standard Model. 
\endhead
    At the present moment the Standard Model is a commonly admitted
and to a sufficient extent experimentally confirmed theory describing
the electromagnetic, weak, and strong interactions. Elementary particles
in the Standard Model are represented by matter fields. They include
the lepton fields $\circpsi^a_{111111}[i]$, $\bulpsi^{a\alpha}_{111}[i]$
and the quark fields $\bulpsi^{a1\alpha\beta}[i]$,
$\circpsi^{a1111\beta}[i]$, $\circpsi^{a\beta}_{11}[i]$. The bullet
over a $\psi$-function is the sign of chiral (left) components, while
the circle is the sign of antichiral (right) components (see details
in \mycite{1}). Through $a=1,\,\ldots,\,4$ we denote a spinor index;
$\beta=1,\,\ldots,\,3$ is a color index, it is peculiar to quark
wave functions only. The index $\alpha=1,\,2$ is a doublet index,
$\psi$-functions possessing this index form doublets. Other
$\psi$-functions are singlets. The index $i=1,\,\ldots,\,3$ is enclosed
into square brackets. It enumerates three generations of leptons and 
three generations of quarks. And finally, each of the above 
$\psi$-functions has some definite number of indices taking the only 
value $\gamma=1$.\par
     Apart from lepton and quark fields, there are gauge fields in the
Standard Model. They are introduced as connection components in covariant
derivatives. For instance, we have the following formula:
$$
\hskip -2em
\aligned
\nabla_{\!q}&\bulpsi^{a1\alpha\beta}[i]
=\nabla_{\!q}[vac]\bulpsi^{a1\alpha\beta}[i]
-\frac{i\,e\,g_1}{\hbar\,c}\,\uCA^1_{q1}
\,\bulpsi^{a1\alpha\beta}[i]\,-\\
&-\,\frac{i\,e\,g_2}{\hbar\,c}\sum^2_{\theta=1}
\uuCA^\alpha_{q\kern 0.4pt\theta}\,
\bulpsi^{a1\theta\kern 0.4pt\beta}[i]
-\frac{i\,e\,g_3}{\hbar\,c}\sum^3_{\theta=1}
\uuuCA^\beta_{q\kern 0.4pt\theta}
\,\bulpsi^{a1\alpha\kern 0.4pt\theta}[i].
\endaligned
\mytag{1.1}
$$
The quantities $\uuuCA^\beta_{q\kern 0.4pt\theta}$ correspond to the
gluon field. The quantities $\uCA^1_{q1}$ and $\uuCA^\alpha_{q\kern 
0.4pt\theta}$ correspond to the electromagnetic and weak fields 
respectively. Upon applying the Higgs mechanism they mix and form
the $4$-potential of the electromagnetic field $A_q$ and the fields of
massive bosons $Z_q$ and $W^{\scriptscriptstyle\pm}_{q111111}$.
Here $q=0,\,\ldots,\,3$ is a covectorial index. The nature of the
index $\theta$ in the formula \mythetag{1.1} is clear from the 
formula itself. It is sufficient to look at the ranges of this index
in sums. Through $g_1$, $g_2$, and $g_3$ we denote three purely
numeric constants, they are parameters of the Standard Model.
And finally, $e$, $\hbar$, and $c$ in the formula \mythetag{1.1}
are the charge of electron, the Planck constant, and the light
speed respectively:
$$
\allowdisplaybreaks
\align
&\hskip -2em
e\approx 4.80420440\cdot 10^{-10}
\,\text{\it g}^{\,\sssize 1/2}\cdot
\text{\it cm}^{\,\sssize 3/2}\cdot
\text{\it sec}^{\sssize\,-1},\\
\vspace{1ex}
&\hskip -2em
\hbar\approx 1.05457168\cdot 10^{-27}
\,\text{\it g}\cdot
\text{\it cm}^{\,\sssize 2}
\cdot\text{\it sec}^{\sssize\,-1},\\
\vspace{1ex}
&\hskip -2em
c\approx 2.99792458\cdot 10^{10}
\,\text{\it cm}\cdot\text{\it sec}^{\sssize\,-1}.
\endalign
$$
The above data are taken from the site 
\myhref{http://physics.nist.gov/cuu/Constants/}{http:/\kern -2pt
/physics.nist.gov/cuu/Constants} of the US National Institute of 
Standards and Technology.\par
    The Higgs field $\varphi^{\kern 0.4pt\alpha111}\kern -1.7pt$ is
the most mysterious field of the Standard Model. It is associated
with an elementary particle (the Higgs boson) which is not detected
experimentally. This fact gives an opportunity for various Higgsless
models, i\.\,e\. theories explaining the absence or invisibility of
Higgs boson in collider experiments. Looking through the review 
\mycite{2} and some recent papers \myciterange{3}{3}{\,--\,}{7}, 
I have found that most of these Higgsless models are produced as 
reductions of some higher-dimensional models. However, there is a 
much more simple purely $4$-dimensional approach for eliminating 
the Higgs field from the stage. It is  described below.
\head
2. Composite Higgs fields.
\endhead
    As we have seen in the beginning of the  previous section (see 
also \mycite{8}), the lepton and quark fields in the Standard Model 
are represented by $\psi$-functions with some definite number of 
indices. We can perform tensor products and contractions over some 
pairs of indices of the same nature. For example, we can write
$$
f^{\,\alpha111}[i]=\sum^4_{a=1}\sum^4_{\bar a=1}D_{a\bar a}\,
\overline{\circpsi^{\bar a}_{111111}[i]}\ \bulpsi^{a\alpha}_{111}[i]
\,\uD^{11}\,\uD^{11}\,\uD^{11}\,\uD^{11}\,\uD^{11}\,\uD^{11}.\quad
\mytag{2.1}
$$
In a more general case we can mix the wave functions from different 
generations and write an expression similar to \mythetag{2.1} for them:
$$
f^{\,\alpha111}[ij]=\sum^4_{a=1}\sum^4_{\bar a=1}D_{a\bar a}\,
\overline{\circpsi^{\bar a}_{111111}[i]}\ \bulpsi^{a\alpha}_{111}[j]
\,\uD^{11}\,\uD^{11}\,\uD^{11}\,\uD^{11}\,\uD^{11}\,\uD^{11}.\quad
\mytag{2.2}
$$
Here $D_{a\bar a}$ are the components of the Hermitian metric $\bold D$
in the bundle of Dirac spinors $DM$ over the space-time manifold $M$.
Similarly, $\uD^{11} $ are the contravariant components of the 
Hermitian metric $\bolduD$ in the one-dimensional bundle $\uU M$ (see
\mycite{8}, \mycite{9}, and \mycite{10} for more details).\par
     The functions \mythetag{2.1} and \mythetag{2.2} have the same 
list of indices as the Higgs filed $\varphi^{\kern 0.4pt\alpha111}
\kern -1.7pt$. \pagebreak They are so called composite Higgs fields produced 
from the lepton fields. The quark fields are also capable to produce 
composite Higgs fields:
$$
\hskip -2em
\gathered
\hat F^{\,\alpha111}[ij]
=\sum^4_{a=1}\sum^4_{\bar a=1}\sum^3_{\beta=1}\sum^3_{\bar\beta=1}
\sum^2_{\theta=1}\sum^2_{\bar\alpha=1}
\uud^{\alpha\theta}\,\uuD_{\theta\bar\alpha}\,
D_{a\bar a}\,\uuuD_{\beta\bar\beta}\,\times\\
\vspace{1ex}
\times\,\overline{\bulpsi^{\bar a1\bar\alpha\bar\beta}[i]}\ 
\circpsi^{a1111\beta}[j]\ \uD_{11}.
\endgathered
\mytag{2.3}
$$
This is not the only way for producing composite Higgs fields from
the quark fields. Here is the other formula for other composite 
Higgs fields:
$$
\check F^{\,\alpha111}[ij]
=\sum^4_{a=1}\sum^4_{\bar a=1}\sum^3_{\beta=1}\sum^3_{\bar\beta=1}
D_{a\bar a}\,\uuuD_{\beta\bar\beta}\ \overline{\circpsi^{\bar a
\bar\beta}_{11}[i]}\ \bulpsi^{a1\alpha\beta}[j]\ \uD^{11}
\,\uD^{11}.\quad
\mytag{2.4}
$$
In addition to $\bold D$ and $\bolduD$, in \mythetag{2.3} and
\mythetag{2.4} we have the components $\uuuD_{\beta\bar\beta}$
of the Hermitian metric $\bolduuuD$ in three-dimensional color-bundle
$S\uuuU M$ over the space-time manifold $M$, the components
$\uuD_{\theta\bar\alpha}$ of the Hermitian metric $\bolduuD$ in
two-dimensional bundle $S\uuU M$, and the components 
$\uud^{\alpha\theta}$ of the skew-symmetric metric tensor $\bolduud$ 
in the same bundle $S\uuU M$ (see \mycite{10} again).
\par
     Composite Higgs fields \mythetag{2.2}, \mythetag{2.3}, and 
\mythetag{2.4} can be used in order to construct the actual Higgs 
field $\varphi^{\kern 0.4pt\alpha111}\kern -1.7pt$ as a linear
combination:
$$
\hskip -2em
\gathered
\varphi^{\kern 0.4pt\alpha111}\kern -1.2pt=
\sum^3_{i=1}\sum^3_{j=1}C[ij]\,f^{\,\alpha111}[ij]\,+\\
+\sum^3_{i=1}\sum^3_{j=1}\hat C[ij]\,\hat F^{\,\alpha111}[ij]\,
+\sum^3_{i=1}\sum^3_{j=1}\check C[ij]\,\check F^{\,\alpha111}[ij].
\endgathered
\mytag{2.5}
$$
The formula \mythetag{2.5} leads to a theory very similar to the
Standard Model. This theory will be considered in a separate paper.
\head
3. Acknowledgments.
\endhead
    I am grateful to I.~R.~Gabdrakhmanov, S.~G.~Glebov, A.~N.~Khairullin,
O.~M.~Ki\-selev, Yu\.~A.~Kordyukov, T.~A.~Semenova, and B.~I.~Suleymanov 
for encouraging interest to my exercises with the Standard Model. I am
also grateful to A.~Sukachev who communicated me the reference to the 
paper \mycite{2}.\par


\newpage
\Refs
\ref\myrefno{1}\by Sharipov~R.~A.\paper A note on metric connections 
for chiral and Dirac spinors\publ e-print 
\myhref{http://arXiv.org/abs/math/0602359/}{math.DG}
\myhref{http://arXiv.org/abs/math/0602359/}{/0602359}
in Electronic Archive \myEarXivlink
\endref
\ref\myrefno{2}\by  Lillie~B., Terning~J., Grojean~Ch\.,
De~Curtis~S., Dominici~D.\paper Higgsless models\pages 407--428
\inbook Workshop on CP Studies and Non-Standard Higgs Physics,
May 2004 -- December 2005\publ CERN\publaddr Geneva\yr 2006
\moreref see also e-print \myhref{http://arXiv.org/abs/hep-ph/0608079/}
{hep-ph/0608079} in Electronic Archive \myEarXivlink
\endref
\ref\myrefno{3}\by Delgado~A., Falkowski~A.\paper Electroweak 
observables in a general 5D background\publ e-print 
\myhref{http://arXiv.org/abs/hep-ph/0702234/}{hep-}
\myhref{http://arXiv.org/abs/hep-ph/0702234/}{ph/0702234}
in Electronic Archive \myEarXivlink
\endref
\ref\myrefno{4}\by Chivukula~R.~S., Simmons~E.~H., Matsuzaki~Sh\.,
Tanabashi~M.\paper The three site model at one-loop\publ e-print 
\myhref{http://arXiv.org/abs/hep-ph/0702218/}{hep-ph/0702218}
in Electronic Archive \myEarXivlink
\endref
\ref\myrefno{5}\by Carena~M., Ponton~E., Santiago~J., 
Wagner~C.~E.~M.\paper Electroweak constraints on warped models 
with custodial symmetry\publ e-print 
\myhref{http://arXiv.org/abs/hep-ph/0701055/}{hep-ph/0701055}
in Electronic Archive \myEarXivlink
\endref
\ref\myrefno{6}\by Hirn~J., Sanz~V.\paper The fifth dimension 
as an analogue computer for strong interactions at the LHC
\publ e-print 
\myhref{http://arXiv.org/abs/hep-ph/0612239/}{hep-ph/0612239}
in Electronic Archive \myEarXivlink
\endref
\ref\myrefno{7}\by Coleppa~B., Di\,Chiara~S., Foadi~R.\paper 
One-loop corrections to the $\rho$ parameter in Higgsless
models\publ e-print 
\myhref{http://arXiv.org/abs/hep-ph/0612213/}{hep-ph//0612213}
in Electronic Archive \myEarXivlink
\endref
\ref\myrefno{8}\by Sharipov R. A.\paper A note on the Standard Model 
in a gravitation field\jour
\myhref{http://arxiv.org/abs/math.DG/0605709}{math.DG/0605709}
in Electronic archive \myEarXivlink
\endref
\ref\myrefno{9}\by Sharipov R. A.\paper A note on Dirac spinors in 
a non-flat space-time of ge\-neral relativity\jour
\myhref{http://arxiv.org/abs/math.DG/0601262}{math.DG/0601262}
in Electronic archive \myEarXivlink
\endref
\ref\myrefno{10}\by Sharipov~R.~A.\paper The electro-weak and color 
bundles for the Standard Model in a gravitation field\publ e-print 
\myhref{http://arXiv.org/abs/math/0603611/}{math.DG/0603611} 
in Electronic Archive \myEarXivlink
\endref
\endRefs
\enddocument
\end